\documentclass[aps,prb,twocolumn, showpacs]{revtex4}

\usepackage{amssymb}
\usepackage{amsmath}
\usepackage{graphicx}
\usepackage{color}

\def\be{\begin{equation}}
\def\ee{\end{equation}}
\def\bea{\begin{eqnarray}}
\def\eea{\end{eqnarray}}

\begin{document}

\title{Giant spin rotation under quasiparticle-photoelectron conversion:\\
Joint effect of sublattice interference and spin-orbit coupling}
\author{Ferdinand Kuemmeth$^1$ and Emmanuel I. Rashba$^{1,2,3}$}
\affiliation{
$^1$Department of Physics, Harvard University, Cambridge, Massachusetts 02138, USA\\ 
$^{2}$Center for Nanoscale Systems, Harvard University, Cambridge, Massachusetts 02138, USA, and\\ 
$^3$Department of Physics, Loughborough University, Leicestershire LE11 3TU, UK}
\date{October 7, 2009}


\begin{abstract}
Spin- and angular-resolved photoemission spectroscopy is a basic experimental tool for unveiling spin polarization of electron eigenstates in crystals. We prove, by using spin-orbit coupled graphene as a model, that photoconversion of a quasiparticle inside a crystal into a photoelectron can be accompanied with a dramatic change in its spin polarization, up to a total spin flip. This phenomenon is typical of quasiparticles residing away from the Brillouin zone center and described by higher rank spinors, and results in exotic patterns in the angular distribution of photoelectrons.
\end{abstract}

\maketitle

Establishing spin polarization of quasiparticles in crystals is of crucial importance for semiconductor spintronics, physics of metallic surfaces, and the emerging field of topological insulators. The latter typically comprises narrow-gap systems with strong spin-orbit coupling. Graphene, a zero-gap conductor, attracts attention due to its linear Dirac-Weyl energy spectrum and prospects for applications. Its quasirelativistic spectrum, manifesting itself in unconventional quantum Hall effect \cite{Novos05,Zhang05} and Klein tunneling \cite{Kats06}, is essentially a nonrelativistic phenomenon originating from two equivalent sublattices, $A$ and $B$; their effect is conveniently accounted for by pseudospin. However, the relativistic effects that entangle their spin and pseudospin degrees of freedom \cite{KM05} lift the spin degeneracy of the energy spectrum. Intrinsic spin-orbit interaction in graphene \cite{DresDresSO} is weak, not exceeding tens of $\mu$eV \cite{Min06,Yao07,BoetTri,Gmitra}, but breaking the up-down symmetry by a substrate can result in a substantial extrinsic spin-orbit coupling. Such a coupling of the scale of 10 meV was discovered by Varykhalov {\it et al.} \cite{VarAuNi} by spin- and angular-resolved photoemission spectroscopy (SARPES) techniques. Despite its modest magnitude, this coupling modifies essentially the zero-gap nonrelativistic spectrum. This makes spin-orbit coupled graphene an excellent platform for unveiling nontrivial effects of spin-orbit coupling on SARPES spectra.

\begin{figure}[h!]
\center \label{figure1}
\includegraphics[width=8.6cm]{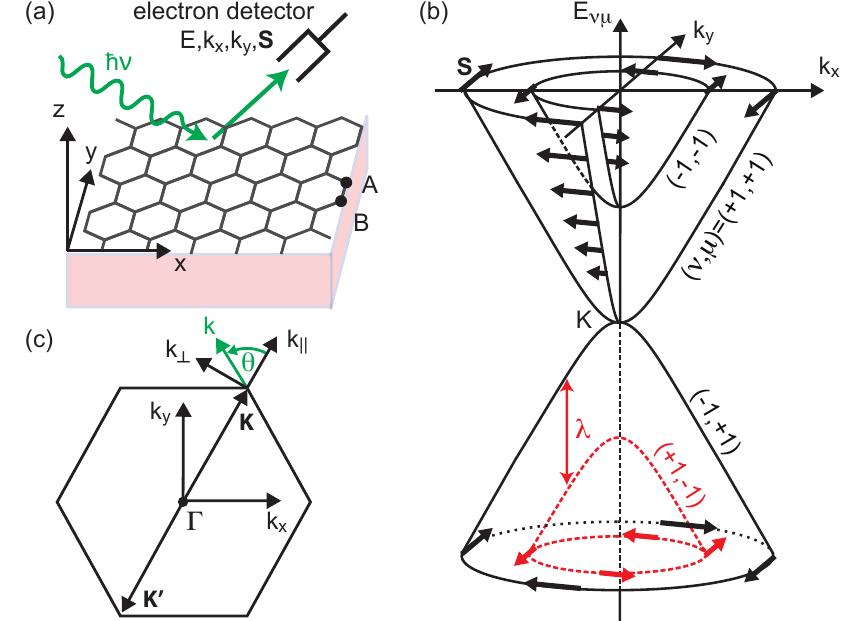}
\caption{
Energy spectrum and spin polarization of quasiparticles in spin-orbit coupled graphene. (a) Schematic of a graphene flake with substrate and a SARPES detector (sensitive to energy, momentum, and spin of photoelectrons). Symmetry between A and B sublattices is preserved. 
(b) Energy spectrum and spin polarization near the $K$ point. Spins are in-plane polarized transverse to the momentum {\bf k}. The magnitude of the spin polarization of quasiparticles (arrows) is proportional to the group velocity. Its sign is shown for $\lambda>0$; it is opposite for $\lambda<0$. (c) Brillouin zone and coordinate systems in the momentum space.
}
\end{figure}

In graphene the zeros of the gap are achieved in two nonequivalent corners of the Brillouin zone (BZ), $K$ and $K^\prime$ [Fig.~1c]. In the vicinity of these points the nonrelativistic quasiparticles (electrons and holes) possess a quasirelativistic energy spectrum $E({\bf k})=\pm\gamma k$, $\bf k$ being a quasimomentum counted relative to the $K (K^\prime)$ point \cite{CastNeto09}. Near the $K$ point, the nonrelativistic Hamiltonian is ${\cal H}_0=\gamma (\mbox{\boldmath$\sigma$}\cdot{\bf k})$, where the pseudospin $\mbox{\boldmath$\sigma$}=(\sigma_1,\sigma_2)$ is a vector of Pauli matrices acting on $(A, B)$ sublattices [Fig.~1(a)]. The leading term in the extrinsic spin-orbit coupling is  ${\cal H}_{\rm so}={1\over2}\lambda(\mbox{\boldmath$\sigma$}\times{\bf s})_z$, where ${\bf s}$ are Pauli matrices of the real spin and $\lambda$ is the spin-orbit coupling constant \cite{KM05,Zarea}; it couples spin to the pseudospin. We disregard the breaking of the $(A,B)$ symmetry by the staggering potential of the substrate; for some substrates it is weak and graphene behaves as quasifreestanding \cite{VarAuNi,Bostwick07}. The two-sublattice structure of graphene is known to result in an interference effect in the nonrelativistic ARPES spectrum that produces strong photoemission anisotropy but does not distort the shape of the isoenergy surfaces \cite{Shirley}. In relativistic spectra, the joint effect of interference and the spin-pseudospin entanglement produces giant rotations of electron spins during photoemission resulting in drastic differences in the spin polarization of quasiparticles inside the crystal and photoelectrons in vacuum. This phenomenon is the focus of this paper.

The Hamiltonian ${\cal H}={\cal H}_0+{\cal H}_{\rm so}$ acts in the space of four-spinors defined in the basis $\{|A\uparrow\rangle,|A\downarrow\rangle,|B\uparrow\rangle,|B\downarrow\rangle\}$, the products of the two-fold degenerate Bloch functions of the $K$ point at $A(B)$ sublattices and spin-up(down) spinors. It is convenient to change to a basis 
$\{|A\uparrow\rangle,|B\downarrow\rangle,|B\uparrow\rangle,|A\downarrow\rangle\}$ and express each bispinor, $\Psi_{\nu\mu}$, in terms of two spinors, $\varphi_{\nu\mu}$ and $\phi_{\nu\mu}$, as $\Psi_{\nu\mu}=\left(\begin{array}{c}\varphi_{\nu\mu}\\\phi_{\nu\mu}\end{array}\right)$; for details see Ref.~\onlinecite{ER09}. The eigenvalues are
\be
E_{\nu\mu}(k)={{\nu\mu}\over{2}}\left(\sqrt{\lambda^2+4\gamma^2k^2}-\mu\lambda\right),
\label{eq1}
\ee
where $\nu,\mu=\pm1$. The spectrum consists of two ungapped and two gapped hyperbolas shifted by $\lambda$ that are shown in Fig.~1(b) with their quantum numbers. The spectrum is similar to  unbiased bilayer graphene \cite{CastNeto09}, but with different nature of eigenstates and narrower gap. In the new basis, matrices of the quasiparticle spin are ${\hat{\bf S}}=\sigma_1{\bf s}$, and their mean values in the $(\nu,\mu)$ eigenstates are 
\be
\langle{\hat{\bf S}}\rangle_{\nu\mu}({\bf k})={{2\mu\gamma({\bf k}\times{\hat{\bf z}})}\over{\sqrt{\lambda^2+4\gamma^2k^2}}}\,;
\label{eq2}
\ee
${\hat{\bf z}}$ being a unit vector perpendicular to the substrate. We note in passing that $\langle{\hat{\bf S}}\rangle$ is proportional to the group velocity $\partial E_{\nu\mu}/\hbar\partial{\bf k}$. The sign of the chirality of eigenspinors $(\varphi_{\nu\mu},\phi_{\nu\mu})$ is determined by $\nu$, the sign of spin polarization by $\mu$, and the sign of the energy by the product $\nu\mu$. The polarization of all branches is depicted in Fig.~1(b). Such a spin-polarized spectrum with $|\lambda|\approx 13$ meV was deduced from SARPES data taken from graphene on a Au/Ni(111) substrate \cite{VarAuNi} for $|E_{\nu\mu}(k)|\agt|\lambda|$; data for lesser energies are not available.

According to Eq. (\ref{eq2}), quasiparticles are in-plane polarized perpendicular to the momentum $\bf k$, and the magnitudes of their spins depend on $k$ and can even vanish. This is well compatible with the fact that ${\hat{\bf S}}^2=3$, because spin-orbit coupling results in large nondiagonal components of $\hat{\bf S}$ in the $2\times2$ spin-pseudospin space, hence restrictions on its diagonal components (``crystal" spin) are relaxed. The scale of the anticipated effect follows from the ratio of ${\cal H}_0$ and ${\cal H}_{\rm so}$ that is about $k/k_\lambda$, where $k_\lambda=|\lambda|/2\gamma$ is a spin-orbit momentum. The term ${\cal H}_0$, Zeeman energy of the pseudospin in an effective magnetic field $\bf k$, describes the $\bf k$-dependence of the interference of Bloch waves scattered by two equivalent sublattices, and this mechanism affects the spin sector through the term ${\cal H}_{\rm so}$. This spin renormalization is strongest for $k/k_\lambda\alt1$ [Fig.~1(b)] and leads to vanishing spin at the symmetry points $K$ and $K^\prime$ as a result of lattice interference. We emphasize that this is impossible for simple lattices, where $\langle{\hat{\bf s}}\rangle^2=1$ is maintained either by in-plane polarization\cite{R60} or by out of plane spin rotation as recently reported \cite{Sakamoto}.

\begin{figure}[h!]
\center \label{figure2}
\includegraphics[width=8.6cm]{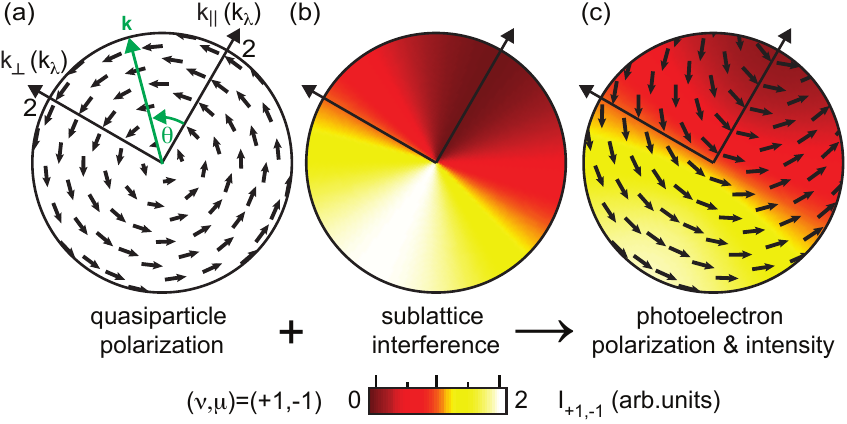}
\caption{
Spin-polarization of quasiparticles and photoelectrons near the $K$ point. (a) Direction (arrows) and magnitude (lengths) of quasiparticle spins of the lower spectrum branch; spins vanish for $k\rightarrow0$. (b) Photoemission probability from non-relativistic graphene ($\lambda=0$). Color indicates the photocurrent $I_{+1,-1}(\theta)$ (dark = 0, bright = 2). (c) Spin polarized photocurrent from the lower spectrum branch; arrows indicate photoelectron spins. Polarization persists at  $k\rightarrow0$ and changes fast near the ``dark corridor" ($k_\parallel>0$, $k_\perp=0$). For $k\leq2$, in units of $k_\lambda$.
}
\end{figure}

While the effect of sublattice interference on the angular dependence of ARPES, unveiled by Shirley {\it et al.} \cite{Shirley}, is well documented for monolayer \cite{VarAuNi,Bostwick07} and bilayer ~\cite{Bostwick07,Mucha08} graphene and for  graphite \cite{Shirley,Matsui02,Zhou06,Leem09}, the recent progress in studying quasiparticles by SARPES techniques, the novelty of results, and the large magnitude of spin-orbit coupling found in a number of systems\cite{VarAuNi,Sakamoto,Henk04,Meier08,VarCavity,Dil08,Hsieh09,Wells09,Meier09,Nishide09,Gierz}, all call for proper understanding of interference-generated spin patterns in photoemission. They are specific for the single-step transformation of Bloch spinors of quasiparticles into photoelectron plane waves with large in-plane (Brillouin) momentum and manifest themselves in singular $\bf k$ dependencies in the $k\rightarrow0$ limit. The Fano effect in atomic photoionization \cite{Fano} and polarization-dependent interband transitions  \cite{Borstel} never show such singularities. They also are absent in the single-step normal photoemission showing remarkable polarization properties depending on the incidence conditions \cite{Feder}.

For calculating photoemission, one needs to employ detailed wave functions that are products of the components of envelope spinors $\Psi_{\nu\mu}$ and Bloch basis functions. At the $K$ point of the BZ [Fig.~1(c)], the $|B\rangle$ functions differ from the symmetrically equivalent to them $|A\rangle$ functions by the phase factor $(-\omega)$, with $\omega=\exp{(i2\pi/3)}$ \cite{Ando00,HHGA}. The $\omega$ factor reflects the effect of sublattice interference on photoemission in terms of the envelope functions, while the specific form of $|A\rangle$ is of no importance as long as small spin-orbit corrections depending on atomic form-factors \cite{Henk04} factorize out and can be disregarded. For nonrelativistic electrons ($\lambda=0$) the interference factor in the photoemission intensity reduces to $\left[(k_-/k)(E/|E|)-\omega\right]$ near the $K$ point, $k\ll K$; here $k_-=k_x-ik_y$. The first term in the brackets originates from ${\cal{H}}_0$, and the second from the interference of outgoing photoelectrons ($\exp\left[i{\bf K}\cdot({\bf R}_A-{\bf R}_B)\right]=\omega$, with ${\bf R}_{A,B}$ for sublattice coordinates). The interference factor is related to the quasiparticle pseudospin whose mean value equals $\langle\sigma_x-i\sigma_y\rangle=-\omega(E/|E|)$ for a quasiparticle with a momentum $\bf{k}$ along ${\bf K}$. The resulting photoemission is highly anisotropic as described by the large $k$ limit of Eq.~(\ref{eq5}) and displayed in Fig. 2(b). 

\begin{figure}[h!]
\center \label{figure3}
\includegraphics[width=8.6cm]{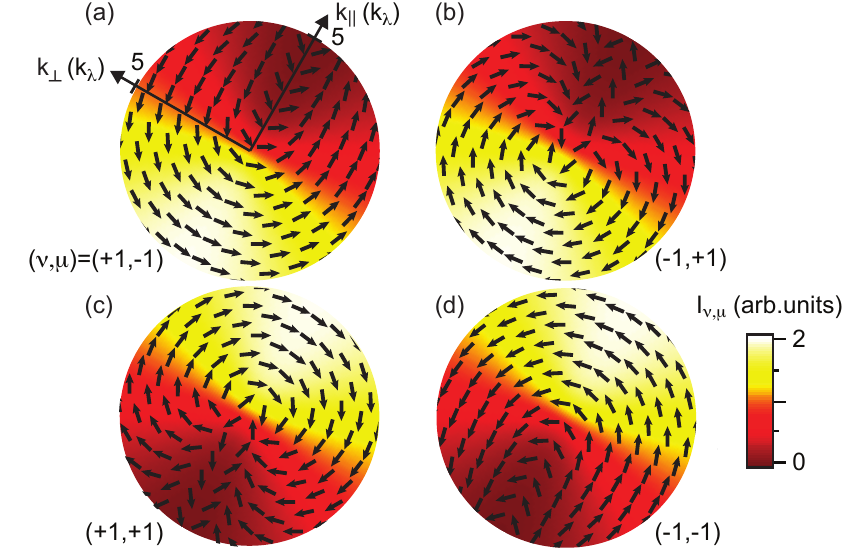}
\caption{
Intensity (colorscale) and spin-polarization (arrows) for photoelectrons emitted from all four branches near the $K$ point, for $k\leq5$ and $\lambda>0$; (a) and (d) for gapped, and (b) and (c) for ungapped branches.  Spins of photoelectrons and quasiparticles become identical only for $k\gg1$ and away from the ``dark corridor". $k$ in units of $k_\lambda$.
}
\end{figure}

For relativistic electrons ($\lambda\neq0$) the envelope spinor $\Phi_{\nu\mu}({\bf k})$, describing the flux of spin-polarized photoelectrons from the $(\nu,\mu)$ eigenstate, can be found by adding the components of $\Psi_{\nu\mu}({\bf k})$ corresponding to the same spin polarizations of the basis spinors, supplied with the proper phase factors originating from $|A\rangle$ and $|B\rangle$ functions
\be
\Phi_{\nu\mu}({\bf k})\propto
\left(\begin{array}{c}
\varphi^{(1)}_{\nu\mu}({\bf k})-\omega\phi^{(1)}_{\nu\mu}({\bf k})\\
\phi^{(2)}_{\nu\mu}({\bf k})-\omega\varphi^{(2)}_{\nu\mu}({\bf k})
\end{array}\right),
\label{eq3}
\ee
where superscripts indicate the upper and lower components of the spinors $\varphi$ and $\phi$. This spinor, defined with the accuracy to a factor depending on the atomic matrix element, intensity of the source, etc., describes the dependence of the spin flux on the azimuth $\theta$ [Fig.~1(c)].

The final form of $\Phi_{\nu\mu}({\bf k})$ is determined by the explicit form of the quasiparticle spinors
\bea
\varphi_{\nu\mu}({\bf k})&=&{{\gamma k}\over{\sqrt{2|E_{\nu\mu}|\sqrt{\lambda^2+4\gamma^2k^2}}}}\left(\begin{array}{c}i\nu k_-^2/k^2\\1\end{array}\right),\,\nonumber\\
\phi_{\nu\mu}({\bf k})&=&{{(k_-/k)E_{\nu\mu}}\over{\sqrt{2|E_{\nu\mu}|\sqrt{\lambda^2+4\gamma^2k^2}}}}\left(\begin{array}{c}i\nu \\1\end{array}\right),
\label{eq4}
\eea
normalized as $\langle\Psi_{\nu\mu}({\bf k})|\Psi_{\nu\mu}({\bf k})\rangle=1$. In particular, the total flux equals $I_{\nu\mu}(k,\theta)\equiv\langle\Phi_{\nu\mu}({\bf k})|\Phi_{\nu\mu}({\bf k})\rangle$
\be
I_{\nu\mu}(k,\theta)=1+2\nu\mu\gamma k\cos\theta/\sqrt{\lambda^2+4\gamma^2k^2}\,.
\label{eq5}
\ee
In the large $k$ limit, $k\gg k_\lambda$, it coincides with the well known result \cite{Shirley,Bostwick07}, $1+\nu\mu\cos\theta$, with its strong $\theta$ dependence originating from the sublattice interference. 
On the contrary, in the small $k$ region, $k\ll k_\lambda$, emission is isotropic due to a strong spin-pseudospin entanglement. 

Our focus is on photoelectron spins. They are in-plane polarized, and in a reference system $(k_\parallel,k_\perp)$ related to the $K$ point [Fig.~1(c)]:
\be
s^{\parallel}_{\nu\mu}(k,\theta)=2\mu\gamma k{{1+\gamma k\cos\theta/E_{\nu\mu}(k)}\over{\sqrt{\lambda^2+4\gamma^2k^2}+2\nu\mu\gamma k\cos\theta}}\sin\theta,
\label{eq7}
\ee
\be
s^\perp_{\nu\mu}(k,\theta)=-2\mu\gamma k{{(1+\gamma k\cos\theta/E_{\nu\mu})\cos\theta-\nu\lambda/2\gamma k}\over{\sqrt{\lambda^2+4\gamma^2k^2}+2\nu\mu\gamma k\cos\theta}}\,.
\label{eq8}
\ee
Figures 2(a) and 2(c) show unidirectional flow for photoelectrons (left to right) while for quasiparticles $s^\perp_{1,-1}(\pi/2)=0$.

Alternatively, spins can be specified by their transverse and longitudinal components, $s^t$ and $s^\ell$, defined as mean values of $({\bf s}\times{\hat{\bf k}})_z$ and $({\bf s}\cdot{\hat{\bf k}})$, respectively; ${\hat{\bf k}}={\bf k}/k$. Then
\be
s^\ell_{\nu\mu}({\bf k})={{\nu\mu\lambda\sin\theta}\over{\sqrt{\lambda^2+4\gamma^2k^2}+2\nu\mu\gamma k\cos\theta}}\,,
\label{eq9}
\ee
and $(s^\ell_{\nu\mu})^2+(s^t_{\nu\mu})^2=1$. Spins of photoelectrons acquire a considerable longitudinal component (unless $\theta=0,\pi$) despite the fact that quasiparticles are transverse polarized. It decreases when $k/k_\lambda\rightarrow\infty$, but retains a considerable value in the moderate $k/k_\lambda$ region. The magnitude of $s^\ell$ remains large even in the region of high brightness ($\theta\sim\pi/2$). While for quasiparticles all spins vanish in the $k\rightarrow0$ limit, for photoelectrons they do not, with $s^\ell_{\nu\mu}(k=0)=\nu\mu\lambda\sin\theta/|\lambda|$ and $s^t_{\nu\mu}(k=0)=\nu\cos\theta$. The angular distribution of the total flux, $I_{+1,-1}$, depends on $\lambda$ [Figs.~2(b) and 2(c)].

Spins of photoelectrons coming from different branches are compared in Fig.~3(a-d) in a wider $k$ range. For $k\alt k_\lambda$, the difference between gapped and ungapped branches strikes the eye: for the former spins follow a simple one-sided flow, while for the latter their direction strongly depends on $k$ and $\theta$. For $k\agt k_\lambda$, polarization is transverse only inside the bright sector. While the transverse polarization follows from the $k\rightarrow\infty$ expansions of Eqs.~(\ref{eq7}) and (\ref{eq8}), the convergence is nonuniform, and asymptotic expansions diverge along the ``dark corridor". Near it, the signs of all $s^\perp_{\nu\mu}$ are {\it opposite} for quasiparticles and photoelectrons [Fig.~3(a) and 3(c)]. Near the corridor, spins rotate at the scale of the azimuths of $(1+\nu\mu\cos\theta)\sim(\lambda/2\gamma k)^2$ that grows narrower with increasing $k/k_\lambda$. The intensity is low and vanishes when $k\rightarrow\infty$, but the regions of the anomalous spin polarization are wide enough to be accessible for experiment.

\begin{figure}[h!]
\center \label{figure4}
\includegraphics[width=8.6cm]{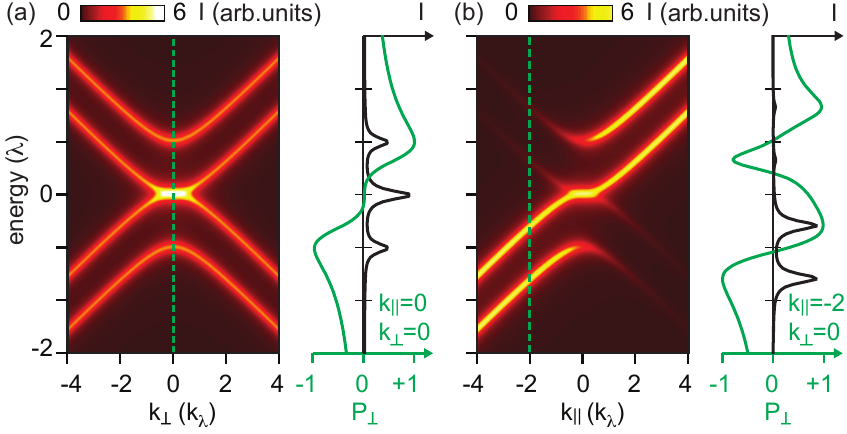}
\caption{
FIG.~4 Energy dependence of the total flux $I_\perp(E)$ and its spin polarization  $P_\perp(E)$ along $k_\perp$. Left panels: energy resolved photocurrent as a function of $k_\perp$ (a) and $k_\parallel$ (b), respectively. Dashed lines indicate the traces of the right panels. Right panels: flux $I_\perp$ (black) and polarization $P_\perp$ (green). See text for details. $E$ in units of $\lambda$, $k$ in units of $k_\lambda$.
}
\end{figure}

Figure~4 displays energy resolved photoemission spectra that can be anticipated. Left panels of (a) and (b)  show Lorentz-widened (width $0.2|\lambda|$) spectra from all spectrum branches in two central sections, {\it vs} $k_\perp/k_{\lambda}$ (with $k_\parallel=0$) and $k_\parallel/k_{\lambda}$ (with $k_\perp=0$).  Due to $\lambda\neq0$, finite intensity persists in panel (b) even inside the ``dark corridor". Right panels exemplify the energy dependence of the total photocurrent $I_\perp(E)=I_P+I_A$ (black) and spin polarization $P_\perp(E)=(I_P-I_A)/I_\perp$ (green), where $I_P$ and $I_A$ are photocurrents polarized parallel and antiparallel to a perfect Mott detector aligned along the positive $k_\perp$ direction. While the details depend on the choice of Lorentzians, the general patterns [such as the sign reversal of $P(E)$ in Fig.~4(a)]  are quite general.

The width of the spectra originating from the lifetime of the holes left after the photoemission is the main obstacle for resolving spin polarized spectra. Because it decreases with decreasing hole energy due to the elimination of scattering channels \cite{Bostwick07}, performing measurements close to the emission threshold should increase the spectral resolution. From this standpoint, small $\lambda$ provides advantages.

For in-plane symmetry preserving substrates, the restrictions imposed by competing mechanisms of SO coupling are essential only within a narrow vicinity of the degeneracy point controlled by intrinsic SO coupling [about 10 $\mu$eV$\sim10^{-3}\lambda$ (Refs. \onlinecite{Min06,Yao07,BoetTri,Gmitra})] and trigonal SO corrections [for $k\alt ak_\lambda^2\ll k_\lambda$, $a$ being a lattice constant, $ak_\lambda\sim10^{-3}$ (Ref.~\onlinecite{Zarea})]; both for the parameter values of Ref.~\onlinecite{VarAuNi}.

It is seen from comparing Figs.~1(b) and 3(a-d) that the difference between the spins of quasiparticles and photoelectrons ($\langle{\hat{\bf S}}\rangle_{\nu\mu}({\bf k})$ and  ${\bf s}_{\nu\mu}(k,\theta)$) differs between three $2\pi/3$ sectors around the $K$ point; for panels (a) and (b) of Fig.~3 it is minimal inside the internal sector that is bright. These data allowed reconstructing the general pattern of spin polarization in the $k\agt k_\lambda$ range \cite{VarAuNi}. Unveiling the spin polarization in the $k\alt k_\lambda$ range is more challenging. Actually, it is an open question whether reconstructing $\langle{\hat{\bf S}}\rangle_{\nu\mu}({\bf k})$, related to four-component spinors $\Psi_{\nu\mu}({\bf k})$, from ${\bf s}_{\nu\mu}(k,\theta)$, related to two-component spinors $\Phi_{\nu\mu}({\bf k})$, is a well posed mathematical problem, and how much spin information is lost during photoemission. In any case, reconstruction of the $k\alt k_\lambda$ region should essentially include the data from external sectors. 

All calculations were performed for spin-orbit coupled graphene. However, the qualitative conclusions are general and are based on (i) a narrow-gap spectrum of quasiparticles with strong spin-orbit coupling described by four-component spinors, and the presence of the $\omega$ phase factors that requires (ii) existence of (nearly) equivalent sublattices and (iii) residence of quasiparticles away from the BZ center. These requirements are fulfilled for the Bi$_{1-x}$Sb$_x$ \cite{Hsieh09,Nishide09} and Pb$_{1-x}$Sn$_x$Te \cite{PV91} compounds. Therefore, we expect that their SARPES spectra of bulk electrons and the electrons from their Tamm-Shockley surface bands should manifest similar anomalies.

In conclusion, we demonstrated a dramatic effect of the spin-pseudospin entanglement on the SARPES spectra of spin-orbit coupled graphene. Similar effects are predicted for different narrow-gap spin-orbit coupled materials.

We thank H. Churchill, J. Williams, C. Marcus and O. Rader for discussions, and acknowledge support in part by NSF, DARPA,  DOD, and Harvard Center for Nanoscale Systems.


\begin{thebibliography}{99}
\bibitem{Novos05}  K. S. Novoselov {\it et al.}, {\it Nature} {\bf 438}, 197 (2005).
\bibitem{Zhang05} Y. Zhang, J. W. Tan, H. L. Stormer, H. L. \& P. Kim, {\it Nature} {\bf 438}, 201 (2005).
\bibitem{Kats06} M. I. Katsnelson, K. S. Novoselov, \& A. K. Geim, {\it Nature Physics} {\bf 2}, 620 (2006).  
\bibitem{KM05} C. L. Kane and E. J. Mele, {\it Phys. Rev. Lett.} {\bf 95}, 226801 (2005).
\bibitem{DresDresSO} G. Dresselhaus and M. S. Dressehaus, {\it Phys. Rev.} {\bf 140}, A401 (1965).
\bibitem{Min06} H. Min, J. E. Hill, N. A. Sinitsyn, B. R. Sahu, L. Kleinman, and A. H. MacDonald, {\it Phys. Rev. B} {\bf 74}, 165310 (2006).
\bibitem{Yao07} Y. Yao, F. Ye, X.-L. Qi, S.-C. Zhang, and Z. Fang, Phys. Rev. B {\bf 75}, 041401(R) (2007).
\bibitem{BoetTri} J. C. Boettger and S. B. Trickey, Phys. Rev. B {\bf 75}, 121402(R) (2007)
\bibitem{Gmitra} M. Gmitra, S. Konschuh, C. Ertler, C. Ambrosch-Draxl, and J. Fabian, arXiv:0904.3315.
\bibitem{VarAuNi} A. Varykhalov, J. S\'anchez-Barriga, A. M. Shikin, C. Biswas, E. Vescovo, A. Rybkin, D. Marchenko, and O. Rader, {\it Phys. Rev. Lett.} {\bf 101}, 157601 (2008).
\bibitem{CastNeto09} A. H. Castro Neto, F. Guinea, N. M. R. Peres, K. S. Novoselov, and A. K. Geim, {\it Rev. Mod. Phys.} {\bf 81}, 109 (2009). 
\bibitem{Zarea} We omit $k$-linear corrections to this term, M. Zarea and N. Sandler, Phys. Rev. B {\bf 79}, 165442 (2009)
\bibitem{Bostwick07} A. Bostwick, T. Ohta, J. L. McChesney, K. V. Emtsev, T. Seyller, K. Horn, E. Rotenberg, {\it New J. Phys.} {\bf 9}, 385 (2007).
\bibitem{Shirley} E. L. Shirley, L. J. Terminello, A. Santoni, and F. J. Himpsel, {\it Phys. Rev. B} {\bf 51}, 13614 (1995).
\bibitem{ER09} E. I. Rashba, {\it Phys. Rev. B} {\bf 79}, 161409(R) (2009).
\bibitem{R60} E. I. Rashba, {\it Sov. Phys. Solid State} {\bf 2}, 1109 (1960). 
\bibitem{Sakamoto} K. Sakamoto, T. Oda, A. Kimura, K. Miyamoto, M. Tsujikawa, A. Imai, N. Ueno, H. Namatame, M. Taniguchi, P. E. J. Eriksson, R. I. G. Uhrberg, {\it Phys. Rev. Lett.} {\bf 102}, 096805 (2009).
\bibitem{Mucha08} M. Mucha-Kruczy\'nski, O. Tsyplyatyev, A. Grishin, E. McCann, V. I. Fal'ko, A. Bostwick, and E. Rotenberg, {\it Phys. Rev. B} {\bf 77}, 195403 (2008).
\bibitem{Matsui02} F. Matsui, Y. Hori, H. Miyata, N. Suganuma, H. Daimon, H. Totsuka, K. Ogawa, T. Furukubo, H. Namba, {\it Appl. Phys. Lett.} {\bf 81}, 2556  (2002).
\bibitem{Zhou06} S. Y. Zhou, G.-H. Gweon, J. Graf, A. V. Fedorov, C. D. Spataru, R. D. Diehl, Y. Kopelevich, D.-H. Lee, S. G. Louie, A. Lanzara, {\it Nature Physics} {\bf 2}, 595 (2006).
\bibitem{Leem09} C. S. Leem, C. Kim, S. R. Park, M.-K. Kim, H. J. Choi, C. Kim, B. J. Kim, S. Johnston, T. Devereaux, T. Ohta, A. Bostwick, E. Rotenberg, Phys. Rev. B {\bf 79}, 125438 (2009).
\bibitem{Henk04} J. Henk, M. Hoesch, J. Osterwalder, A. Ernst, and P. Bruno, {\it J. Phys.: Condens. Matter} {\bf 16}, 7581 (2004).
\bibitem{Meier08} F. Meier, H. Dil, J. Lobo-Checa, L. Patthey, J. Osterwalder,  {\it Phys. Rev. B} {\bf 77}, 165431 (2008).
\bibitem{VarCavity} A. Varykhalov, J. S\'anchez-Barriga, A. M. Shikin, W. Gudat, W. Eberhardt, O. Rader, {\it Phys. Rev. Lett.} {\bf 101}, 256601 (2008).
\bibitem{Dil08} J. H. Dil, F. Meier, J. Lobo-Checa, L. Patthey, G. Bihlmayer, and J. Osterwalder, {\it Phys. Rev. Lett.} {\bf 101}, 266802 (2008).
\bibitem{Hsieh09} D. Hsieh, Y. Xia, L. Wray, D. Qian, A. Pal, J. Dil, J. Osterwalder, F. Meier, G. Bihlmayer, C. Kane, Y. Hor, R. Cava, M. Hasan, {\it Science} {\bf 323}, 919 (2009).
\bibitem{Wells09} J. W. Wells, J. H. Dil, F. Meier, J. Lobo-Checa, V. N. Petrov, J. Osterwalder, M. M. Ugeda, I. Fernandez-Torrente, J. I. Pascual, E. D. L. Rienks, M. F. Jensen, Ph. Hofmann, {\it Phys. Rev. Lett.} {\bf 102}, 096802 (2009).
\bibitem{Meier09} F. Meier, V. Petrov, S. Guerrero, C. Mudry, L. Patthey, J. Osterwalder, J. H. Dil, Phys. Rev. B {\bf 79}, 241408(R) (2009).
\bibitem{Nishide09} A. Nishide, A. A. Taskin, Y. Takeichi, T. Okuda, A. Kakizaki, T. Hirahara, K. Nakatsuji, F. Komori, Y. Ando, I. Matsuda,  arXiv:0902.2251.
\bibitem{Gierz} I. Gierz, T. Suzuki, E. Frantzeskakis, S. Pons, S. Ostanin, A. Ernst, J. Henk, M. Grioni, K. Kern, and C. R. Ast, Phys. Rev. Lett. {\bf 103}, 046803 (2009). 
\bibitem{Fano} U. Fano, Phys. Rev. {\bf 178}, 131 (1969).
\bibitem{Borstel} G. Borstel and M. W\"{o}hlecke, Phys. Rev. B {\bf 24}, 2321 (1981).
\bibitem{Feder} E. Tamura and R. Feder, Solid State Communic. {\bf 79}, 989 (1991).
\bibitem{Ando00} T. Ando, {\it J. Phys. Soc. Japan} {\bf 69}, 1757 - 1763 (2000).
\bibitem{HHGA} D. Huertas-Hernando, F. Guinea, and A. Brataas, Phys. Rev. B {\bf 74}, 155426 (2006).
\bibitem{PV91} O. A. Pankratov and B. A. Volkov, in: {\it Landau Level Spectroscopy}, North-Holland (Amsterdam, 1991), p. 817

\end{thebibliography}
\end{document}